\documentclass[letterpaper,english,aps,prb,floatfix,showpacs,twocolumn,amsfonts,amssymb,superscriptaddress,longbibliography]{revtex4-2}

\usepackage[english]{babel}

\usepackage{amsmath}
\usepackage{makecell}
\usepackage{graphicx}
\usepackage{hyperref}

\usepackage{balance}
\usepackage[T1]{fontenc}
\usepackage[applemac]{inputenc}
\usepackage[english]{babel}
\usepackage{graphics}
\usepackage{amssymb}
\usepackage{amsmath}
\usepackage{overpic}
\usepackage{bbold}
\usepackage[toc]{appendix}

\usepackage{xcolor}
\usepackage{tabularx}
\usepackage{soul}
\usepackage{commath}
\usepackage{tikz}
\usepackage{physics}

\usepackage{hyperref}

\newcommand{\be}{\begin{equation}}
\newcommand{\ee}{\end{equation}}
\newcommand{\bspl}{\begin{split}}
\newcommand{\espl}{\end{split}}
\newcommand{\bea}{\begin{eqnarray}}
\newcommand{\eea}{\end{eqnarray}}

\newcommand{\bd}{\boldsymbol}

\def\a{\alpha}
\def\b{\beta}
\def\e{\varepsilon}
\def\d{\delta}
\def\g{\gamma}

\def\m{\mu}
\def\l{\lambda}
\def\th{\theta}

\def\n{\nu}

\def\s{\sigma}

\def\D{\Delta}
\def\O{\Omega}

\def\hH{\hat{H}}

\def\hr{\hat{\rho}}

\def\ra{\rightarrow}

\def\pll{\parallel}

\def\pd{\partial}
\def\nb{\nabla}
\def\bdnb{\bd{\nb}}
\def\bk{{\bf k}}
\def\bh{{\bf h}}

\def\br{{\bf r}}

\def\ba{{\bf a}}
\def\bA{{\bf A}}

\def\bE{{\bf E}}

\def\bD{{\bf D}}

\def\bJ{{\bf J}}

\def\bp{{\bf p}}
\def\br{{\bf r}}

\def\bv{{\bf v}}
\def\bx{{\bf x}}
\def\by{{\bf y}}

\def\bo{{\bf 0}}

\def\bxi{{\boldsymbol{\xi}}}
\def\bO{{\boldsymbol{\Omega}}}

\def\nn{\nonumber}
\def\lb{\label}
\def\pref#1{(\ref{#1})}

\definecolor{darkred}{rgb}{0.55, 0.0, 0.0}
\definecolor{darkpowderblue}{rgb}{0.0, 0.2, 0.6}

\begin{document}

\title{Hallmarks of spin textures for high-harmonic generation in two-dimensional materials}

\author{F. Gabriele}
\affiliation{CNR-SPIN, c/o Universit\`a di Salerno, IT-84084 Fisciano (SA), Italy}

\author{C. Ortix}
\affiliation{Dipartimento di Fisica "E. R. Caianiello", Universita` di Salerno, IT-84084 Fisciano (SA), Italy}
\affiliation{CNR-SPIN, c/o Universit\`a di Salerno, IT-84084 Fisciano (SA), Italy}

\author{M. Cuoco}
\affiliation{CNR-SPIN, c/o Universit\`a di Salerno, IT-84084 Fisciano (SA), Italy}

\author{F. Forte}
\affiliation{CNR-SPIN, c/o Universit\`a di Salerno, IT-84084 Fisciano (SA), Italy}

\begin{abstract}
Spin-orbit coupling and quantum geometry are fundamental aspects in modern condensed matter physics, with their primary manifestations in momentum space being spin textures and Berry curvature. In this work, we investigate their interplay with high-harmonic generation (HHG) in two-dimensional non-centrosymmetric materials, with an emphasis on even-order harmonics. Our analysis reveals that the emergence of finite even-order harmonics necessarily requires a broken twofold rotational symmetry in the spin texture, as well as a non-trivial Berry curvature in systems with time-reversal invariance. This symmetry breaking can arise across various degrees of freedom and impact both spin textures and optical response via spin-orbit interactions. We also show that HHG is particularly sensitive to dynamical rotational-symmetry breaking, as even high-order components can be modulated by a time-dependent symmetry breaking. These findings underscore the potential of HHG as a tool for exploring electronic phases with broken rotational symmetry, as well as the associated phase transitions in two-dimensional materials, and provide novel perspectives for designing symmetry-dependent nonlinear optical phenomena.
\end{abstract}

\maketitle

\section{Introduction}

Recent technological advances in laser science have enabled the development of high-intensity light sources across a wide range of energies. Such strong pulses have been increasingly employed in experimental condensed matter physics to induce nonlinear optical responses in crystalline solids. A notable example is high-harmonic generation (HHG), a nonlinear process in which a material exposed to light at a specific frequency emits radiation at multiples of this fundamental frequency\cite{ghimire_natphys11,Hafez2018-eg}. A key distinction in solid-state HHG is that odd-order harmonics are universally present, whereas even-order harmonics are suppressed by inversion symmetry. Surface, interfaces and certain two-dimensional (2D) materials where inversion symmetry is naturally broken offer potential avenues for observing even-order harmonic emission. 2D non-centrosymmetric structures of materials with substantial spin-orbit coupling (SOC)
are generally
characterized by a finite Rashba SOC\cite{Rashba1960,Dresselhaus1955}. This leads to energy band splitting and the emergence of momentum-dependent spin textures -- a variation of the spin direction over the Brillouin zone. At interfaces with trigonal crystal symmetries, these spin textures can also induce a non-zero Berry curvature even if magnetic order is absent\cite{Xiao2010,Mercaldo2023,lesne_natmat23}. 
The presence of SOC and Berry curvature in two dimensions is of significant importance for spintronics\cite{soumyanarayanan_nat16,premasiri_cndmat19,Manchon2015} and topological phenomena\cite{ren_review16}, respectively. Moreover, recent studies have shown that both SOC and Berry curvature are fundamental physical ingredients for HHG: the former governs terahertz third-harmonic generation in transition metals\cite{salikhov2023}, while a direct correspondence between the latter and second-harmonic generation has been established in spin-orbit-free systems\cite{luu_natcomm18,avetissian_prb20,lou_optexp21,yue_prl23}. 

A relevant question is whether and how Berry curvature and SOC cooperate in non-centrosymmetric 2D materials for HHG, particularly in the context of even-order harmonics. Recent studies on the nonlinear Hall effect with time-reversal symmetry\cite{ortix_review21,lesne_natmat23,Sodemann2015,mak24,Ma2019} have established a close connection between SOC, Berry curvature, and second-harmonic generation using a semiclassical approach\cite{Orenstein2021}. The aim of this paper is to demonstrate, using HHG selection rules\cite{lysne_prb20} and explicit calculations, the interplay between SOC, quantum geometry, and high-order harmonics in 2D non-centrosymmetric systems, with a focus on the relation between the emergence of non-zero even-order harmonics and distinct symmetry-allowed spin textures. Note that in certain magnetic materials, non-zero second-order response has been attributed to distinctive magnetic patterns both with\cite{fiebig_optica05,sun_nat19} and without\cite{xiao_npj23,smejkal_natrevmat22} SOC, but without accounting for the contribution of the Berry curvature and the symmetry properties of the spin texture. Our analysis establishes that a spin texture that breaks a $\hat{\mathcal{C}}_2$
twofold rotational symmetry is necessary for observing even-order harmonics. The Berry curvature imposes constraints only under time-reversal symmetric conditions: its non-vanishing is essential for the appearance of even-order harmonics. By contrast, when time-reversal symmetry is broken even-order harmonics can persist even in the complete absence of Berry curvature. Finally, we discuss the case in which twofold rotational symmetry is dynamically broken in time at a certain frequency, leading to amplitude modulation of even high-order harmonics as a function of this driving frequency.

\section{Spin textures and HHG spectrum}

Due to SOC, spin and charge degrees of freedom, and therefore spin textures and high-harmonic response, are interconnected. In particular, both the pattern of the spin texture and high-order harmonics are dictated by the point-group symmetries of the system under investigation: these are unitary operators $\hat{\mathcal{P}}$ that commute with the Hamiltonian $\hat{H}_0$ of the system in the absence of external sources\cite{dresselhaus}. Assuming there is no residual ${\mathcal U}(1)$ spin symmetry -- which is equivalent to stating that the horizontal mirror symmetry is broken -- a 2D system of spin-1/2 fermions can exhibit two distinct types of point-group symmetries. These correspond to proper rotations with respect to the out-of-plane ($z$) axis, and vertical mirrors. Additionally, the system can be equipped with time-reversal symmetry. Here, we examine the impact of these symmetries on both spin textures and high-order harmonics. In particular, HHG is studied in the case of an external light pulse polarized in the in-plane direction, along which the system is space-periodic and the 2D crystal momentum $\mathbf{k} \equiv (k_x, k_y, 0)$ is a good quantum number. The optical response of the system, associated with the current density
$\bJ(t)\equiv\left(J_x(t),J_y(t),0\right)$, can be obtained by incorporating the external vector potential $\bA(t)$, which is related to the electric field $\bE(t)$ via $\bE(t)=-\pd_t\bA(t)$ in the velocity gauge, into $\hat{H}_0$. This is accomplished using the Peierls substitution $\bk\ra\bk+\bA(t)$ (where $e=c=1$ for simplicity)\cite{lysne_prb20}. The details about the computation of $\bJ$ are provided in Appendix A. Our focus is now on the system's general response under a monochromatic driving field $\bA(t)=\Re\lbrace \bA_0\exp(i\O t)\rbrace$ oscillating with frequency $\O$ or period $T=2\pi/\O$, with $\bA_0$ the polarization vector: for linearly-polarized light it is $\bA_0\propto \left(\cos\th,\sin\th,0\right)$, with $\th$ the polarization angle, while $\bA_0\propto (2)^{-1/2}\left(1,\pm i,0\right)$ if light is left/right-hand circularly-polarized. Owing to the time-periodicity of the external probe, the resulting time-dependent Hamiltonian, $\hat{H}(t)$, and the current density, $\bJ(t)$, are also time-periodic. This allows for the Fourier expansion
\be
\bJ(t)=\sum_n \bJ_n e^{in\O t},
\ee
\begin{figure}[t!]
    \centering
    \includegraphics[width=0.45\textwidth,keepaspectratio]{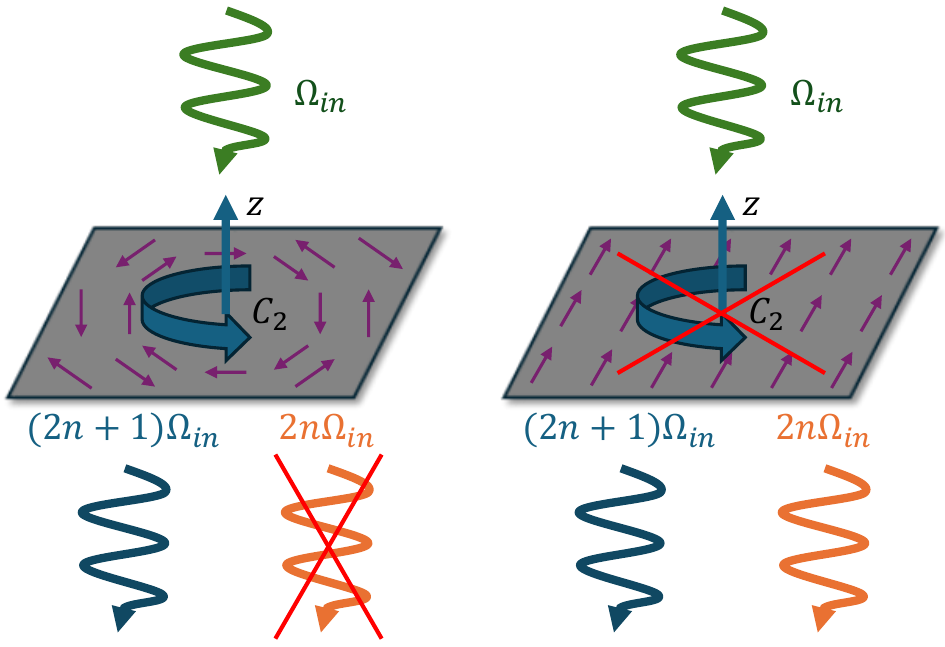}
    \caption{Illustration of spin-texture patterns in momentum space for high-harmonic generation. Left: in 2D systems with spin textures invariant under twofold rotational symmetry, $\hat{\mathcal{C}}_2$, there is no emission of even-order harmonics. Right: for spin textures breaking $\hat{\mathcal{C}}_2$ symmetry, the emission of even-order harmonics is allowed.}
    \lb{fig1}
\end{figure}
where the Fourier component $\bJ_n$ encodes the information on the emission of the $n$-th-order harmonic. The time-dependent Hamiltonian $\hat{H}(t)$ is generally not invariant under the \textit{static} point-group symmetry $\hat{\mathcal{P}}$. As we show in details in Appendix B, one must find a suitable combination of the point-group symmetry and an operator acting on the time variable. This combined symmetry, referred to as \textit{dynamical} symmetry\cite{janssen_phys69,alon_prl98,neufeld_natcomm19,lysne_prb20}, must ensure that both the momentum and the vector potential transform identically. Consider, for instance,
the $\hat{\mathcal{C}}_2$ symmetry operation, which rotates the system by $\pi$ around the $z$ axis, resulting in a sign flip of both momentum and spin. The sign flip of $\mathbf{k}$ can be compensated by applying the time-translation operator $\hat{\mathcal{T}}_{T/2} = \exp\left(i \frac{T}{2} \partial_t\right)$, which shifts time by half a period and thereby reverses the sign of the vector potential. Consequently, the combined operator $\hat{g} \equiv \hat{\mathcal{C}}_2 \otimes \hat{\mathcal{T}}_{T/2}$ serves as a dynamical symmetry of the time-dependent Hamiltonian. In general, a dynamical symmetry $\hat{g}\equiv\hat{\mathcal{P}}\otimes\hat{\mathcal{T}}_{\tau}$ that includes a time translation by a generic interval $\tau$ imposes the following constraint on the $n$-order current
\be
P\bJ_n=e^{i\O\tau}\bJ_n^{(*)},
\lb{currSR}
\ee
where $P$ is the $2\times 2$ matrix associated with the action of $\hat{\mathcal{P}}$ on the two components of the current. Equation \pref{currSR} is valid without or with complex conjugation when $\hat{\mathcal{P}}$ is unitary or anti-unitary, respectively. Its derivation is detailed in Appendix B; here, we focus on its implications.

For $\hat{\mathcal{C}}_2$-symmetric materials illuminated by monochromatic light, whether linearly or circularly polarized, $\tau=T/2$ represents half the period and $P=-\mathbb{1}$ corresponds to minus the 2x2 identity. As a result,
\be
\hat{H}_0
\text{ is }\hat{\mathcal{C}}_2
\text{-invariant}
\quad
\Longrightarrow
\quad
\bJ_n=(-1)^{n+1}\bJ_n.
\lb{C2J}
\ee
Eq.\ \pref{C2J} leads to the crucial consequence that, for even $n$, $\bJ_n$ vanishes. Beside, we recall that, as a consequence of the $\hat{\mathcal{C}}_2$-invariance, the spin texture $\boldsymbol{\s}_\bk\equiv\langle\hat{\boldsymbol{\s}}\rangle_\bk$ transforms as follows:
\be
\hat{H}_0 \text{ is } \hat{\mathcal{C}}_2\text{-invariant}
\Longrightarrow
\sigma_{-\mathbf{k}}^{xy} = -\sigma_{\mathbf{k}}^{xy}, \quad \sigma_{-\mathbf{k}}^{z} = \sigma_{\mathbf{k}}^{z}.
\lb{C2spin}
\ee
Eqs.\ \pref{C2J} and \pref{C2spin}, whose key messages are illustrated in Fig.\ \ref{fig1}, constitute the first central result of this paper: in a 2D non-centrosymmetric system where the spin texture is $\hat{\mathcal{C}}_2$-invariant the emission of even-order harmonics is forbidden. This important criterion implies that systems with $\hat{\mathcal{C}}_2$-broken spin texture can, in principle, emit even-order harmonics. Before discussing this class of materials, we note that the one-to-one correspondence between the HHG selection rules and the spin texture in the presence of SOC holds a more general validity.

To provide another example, we consider systems with a generic rotational invariance $\hat{\mathcal{C}}_N$ with respect to the angle $\phi=2\pi/N$, where $N=2,3,4,6$, under a circularly-polarized pump field. In this case, the dynamical symmetry is $\hat{g}\equiv \hat{\mathcal{C}}_N
\otimes
\hat{\mathcal{T}}_{T/N}$, and as discussed in detail in Appendix B, it imposes the constraint
$Q_N
\bJ_n=
\exp\left(
i 2\pi/N
\right)
\bJ_n$,
where $Q_N$ is the matrix associated with the rotation of $2\pi/N$. Apart from the phase factor arising from the time-dependence of the driven Hamiltonian, the last equation closely resembles the corresponding constraint on the spin texture, which is $Q_N\boldsymbol{\s}_\bk=\boldsymbol{\s}_{Q_N\bk}$. These equations once again highlight the close connection between spin textures and high-order harmonics. 

When considering vertical mirrors, they impose constraints on the polarization of the harmonics. Here, we focus on a representative result for a pump field that is linearly polarized perpendicular to a vertical mirror, while referring to Appendix B for a comprehensive analysis of the entire set of cases. Denoting $\bJ^\pll$ and $\bJ^\perp$ as the components of the optical current parallel and perpendicular, respectively, to the mirror direction, we have
\be
\bJ_n^\parallel = \mathbf{0} \text{ for even } n, \quad \bJ_n^\perp = \mathbf{0} \text{ for odd } n.
\lb{mirror}
\ee
This can be compared to the corresponding transformation of the spin texture, $\boldsymbol{\s}_{\bk_\pll,-\bk_\perp}^\pll=
-\boldsymbol{\s}_{\bk_\pll,
\bk_\perp}^\pll$ and $
\boldsymbol{\s}_{\bk_\pll,-\bk_\perp}^\perp=
\boldsymbol{\s}_{\bk_\pll,
\bk_\perp}^\perp$. 

\textit{Role of the Berry curvature}--- Having discussed how the symmetries of the spin texture affect even-order harmonics, we now turn to the role of Berry curvature. While twofold $\hat{\mathcal{C}}_2$-rotational symmetry remains the main factor governing the emission of even-order harmonics in a 2D non-centrosymmetric system, a vanishing Berry curvature can also suppress the generation of these harmonics in the presence of time-reversal symmetry, even if $\hat{\mathcal{C}}_2$ symmetry is broken\footnote{Such a result holds true for systems with two degrees of freedom, e.g., spin-$1/2$ electrons; for larger numbers of degrees of freedom, the richer algebra generally prevents situations where Berry curvature vanishes if $\hat{\mathcal{C}}_2$-rotational symmetry is broken, apart from a few specific cases where explicit computations show that the result still applies.}. To prove our claim, we examine the current density in the basis of the unperturbed energy bands. The band-resolved density matrix elements, $\rho_{\bk \mu\nu}(t)$, with $\mu$ and $\nu$ denoting band indices, obey, as shown, e.g., in Appendix C or Ref.\ \cite{aversa_prb95}, the Boltzmann equation
\bea
&&\pd_t\rho_{\bk \mu\nu}(t)=
-i(\e_{\bk \mu}-\e_{\bk \nu})
\rho_{\bk \mu\nu}(t)-\bE(t)\cdot\bdnb_\bk
\rho_{\bk \mu\nu}(t)\nn\\
&+&i\bE(t)\cdot\sum_\l
\left(\bxi_{\bk \mu\l}\rho_{\bk \l\nu}(t)-
\rho_{\bk \mu\l}(t)
\bxi_{\bk \l\nu}\right),
\lb{boltz}
\eea
with the initial condition $\rho_{\bk mn}(t\rightarrow-\infty)=f(\e_{\bk m})\d_{mn}$. Here, $\bxi_{\bk\mu\nu}$ is the band-resolved Berry connection. Once Eq.\ \pref{boltz} is solved, the optical current can be computed as (see Appendix C and Refs.\ \cite{aversa_prb95,avetissian_prb20})
\bea
\bJ(t)
=\frac{1}{N}
\sum_{\bk,\mu\nu}\rho_{\bk\mu\nu}(t)
\left(
\bv_{\bk\nu\mu}^{(intra)}
+
\bv_{\bk\nu\mu}^{(inter)}
+
\bv_{\bk\nu\mu}^{(anom)}(t)
\right),\nn\\
\lb{bJ}
\eea
where 
\be
\bv_{\bk\mu\nu}^{(intra)}=\bdnb_\bk\e_{\bk\mu}\d_{\mu\nu},\quad
\bv_{\bk\mu\nu}^{(inter)}=-i(\e_{\bk\mu}-\e_{\bk\nu})\bxi_{\bk\mu\nu}
\ee
are the diagonal intra-band velocity and the inter-band velocity, respectively, while 
\bea
\bv_{\bk\mu\nu}^{(anom)}
=-\bE(t)\times\bO_{\bk\mu\nu}
\eea
is the and anomalous velocity. Here, $\bO_{\bk\mu\nu}\equiv \sum_\l\bD_{\bk\m\l}\times\bxi_{\bk\l\n}$ is the band-resolved Berry curvature, with $\bD_{\bk\m\n}\equiv \d_{\m\n}
\bdnb_\bk-i\bxi_{\bk\m\n}$ the covariant derivative. $\bO_{\bk\mu\mu}$ corresponds to the Berry curvature of the $\mu$-th band. If the Berry curvature of each band vanishes everywhere in momentum space, i.e., $\bO_{\bk\mu\mu} = \mathbf{0}$ for all $\bk$, and the system preserves time-reversal invariance--i.e, $\e_{-\bk\m}=\e_{\bk\m}$ and $\bxi_{-\bk\m\n}=\bxi_{\bk\m\n}^*$--then the emission of even-order harmonics is forbidden. When $\bO_{\bk\mu\mu}=\bo$, indeed, a gauge can always be chosen such that $\bxi_{\bk\mu\mu}=\bo$ and $\bxi_{\bk\mu\nu}$ is purely imaginary for $\mu\neq \nu$. Time-reversal invariance ensures, for a purely imaginary Berry connection, that $\bxi_{-\bk\mu\nu}=-\bxi_{\bk\mu\nu}$ and, consequently, $\bO_{-\bk\mu\nu}=\bO_{\bk\mu\nu}$. It then follows that $\rho_{-\bk\mu\nu}(t+T/2)=\rho_{\bk\mu\nu}(t)$, as they represent the same solution to Eq.\ \pref{boltz}. Given that $\bv_{-\bk\mu\nu}^{(inter/intra)}=-\bv_{\bk\mu\nu}^{(inter/intra)}$ for all $\mu$ and $\nu$, $\bv_{\bk\mu\mu}^{(anom)}(t)=\bo$, and $\bv_{-\bk\mu\nu}^{(anom)}(t+T/2)=-\bv_{\bk\mu\nu}^{(anom)}(t)$ for $\mu\neq\nu$, it follows that $\bJ(t+T/2)=-\bJ(t)$. Therefore, in the presence of time-reversal symmetry,
\be
\bO_{\bk\mu\mu}=\bo
\quad
\Longrightarrow
\quad
\bJ^{(n)}=(-1)^{n+1}\bJ^{(n)}.
\lb{berry}
\ee
This leads to a conclusion similar to that of Eq.\ \pref{C2J}, namely, that in time-reversal invariant systems with zero Berry curvature, $\bJ^{(n)}=\bo$ for even $n$.

\begin{figure*}[t!]
    \centering
    \includegraphics[width=1.0\textwidth,keepaspectratio]{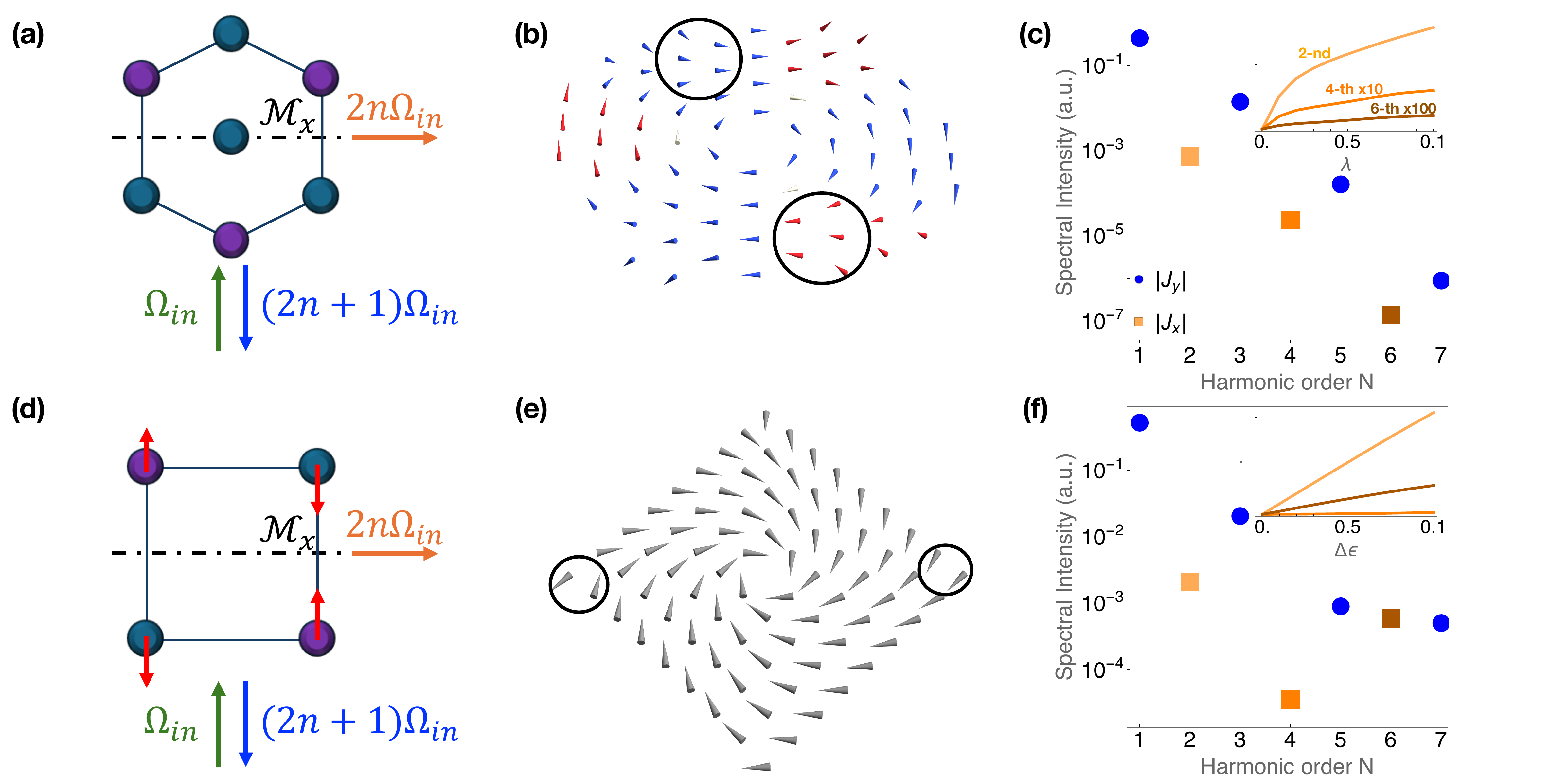}
    \caption{{Spin textures and high-harmonic response for different physical configurations.} Schematic illustration of the trigonal lattice (a) with threefold rotational symmetry and antiferromagnetic pattern (d) with broken twofold rotational symmetry. 
    Both systems possess a vertical mirror symmetry $\mathcal{M}_x$ along the $x$-axis, and in both cases, the incident field (green arrow) is polarized perpendicular to this direction. Blue and orange arrows represent the expected polarizations for odd- and even-order harmonics, respectively. Panels (b) and (e) show the spin textures in momentum space corresponding to the physical configuration in (a) and (d), respectively. The color legend indicates the out-of-plane spin component $\langle \sigma_z \rangle$, with color blue, gray and red corresponding to $\langle \sigma_z \rangle=-1$, $0$ and $1$, respectively. Black circles mark spots in the spin textures where $\hat{\mathcal{C}}_2$ symmetry, as described by Eq.\ \pref{C2spin}, is broken. Panels (c) and (f) show the HHG spectra (in logarithmic scale) for the time-reversal-invariant trigonal configuration and the antiferromagnetic pattern, respectively. The Hamiltonian parameters, in units of the hopping energy $t$, are: $\g_R=0.1$ for both cases; $\l=0.1$ in (c); and $\D\e=0.1$, $B=0.1$ in (f). The computations were performed at temperature $T=0$ with chemical potential $\mu=1.0$. In both cases, the incident field has intensity $|\bA_0|=1.0$ and frequency $\O_{in}=0.5$. Polarizations of odd- (blue dots) and even-order (orange squares, with different shades of orange) harmonics are consistent with the outcomes of Eq.\ \pref{mirror}. The insets show the intensity of even-order harmonics (in linear scale) as functions of the parameters controlling the $\hat{\mathcal{C}}_2$-symmetry breaking: the trigonal warping $\lambda$ in (c) and the charge imbalance $\Delta \epsilon$ in (f). The units and color legend of even-order harmonics in each inset are consistent with those of the corresponding panel.}
    \lb{fig2}
\end{figure*}

\section{HHG in systems with broken $\hat{\mathcal{C}}_2$-symmetry}

Having identified the $\hat{\mathcal{C}}_2$-invariant spin texture as a sufficient condition for the suppression of even-order harmonics, we now examine scenarios where spin textures break this symmetry, enabling their emission. In this context, it is essential to distinguish between the two cases where time-reversal (TR) symmetry is either preserved or broken. 

\textit{Systems with broken $\hat{\mathcal{C}}_2$ and preserved TR symmetry}--- We consider the spin texture and high-harmonic response of a TR-invariant trigonal lattice with threefold rotational symmetry ($\hat{\mathcal{C}}_3$) that explicitly breaks $\hat{\mathcal{C}}_2$ symmetry. The Hamiltonian is given by
\be
\hH_0=\e_\bk\hat{\s}_0+\bh_\bk\cdot\hat{\boldsymbol{\s}},
\label{HC3}
\ee
where $\hat{\s}_0$ is the identity matrix and $\hat{\boldsymbol{\s}}$ represents the Pauli matrices. Here, having defined $\ba_{1/2}=\pm \left(\sqrt{3}/2\right)\hat{\bx}-\left(1/2\right)\hat{\by}$ and $\ba_3=\hat{\by}$, $\e_\bk=-2t\left[
\cos\left(\ba_1\cdot\bk\right)+
\cos\left(\ba_2\cdot\bk\right)+
\cos\left(\ba_3\cdot\bk\right)\right]$ is the tight-binding energy dispersion, $h_\bk^{(x)}=-
\g_R\boldsymbol{\eta}_\bk\cdot\hat{\by}$ and $h_\bk^{(y)}=
\g_R\boldsymbol{\eta}_\bk\cdot\hat{\bx}$, with $\boldsymbol{\eta}_\bk=2\left[
\ba_1\sin\left(\ba_1\cdot\bk\right)+
\ba_2\sin\left(\ba_2\cdot\bk\right)+\ba_3\sin\left(\ba_3\cdot\bk\right)\right]$, describe Rashba spin-momentum locking, and $h_\bk^{(z)}=2\l\left[
\sin\left(\ba_1\cdot\bk\right)+
\sin\left(\ba_2\cdot\bk\right)+\sin\left(\ba_3\cdot\bk\right)\right]$ is the trigonal warping term. While the Rashba coupling is invariant under any rotation, $h_\bk^{(z)}$ breaks $\hat{\mathcal{C}}_2$, reflecting the $\hat{\mathcal{C}}_3$ symmetry of the underlying lattice, and preserves a vertical mirror along the $x$ axis, see Fig.\ \ref{fig2}(a). It also generates a non-zero Berry curvature, $\bO(\bk)\propto\l$\cite{lesne_natmat23}, which is crucial in this case for enabling even-order harmonics. For our calculations, we chose parameters representative of trigonal perovskite oxide materials, such as LAO/STO or LAO/KTO interfaces, where spin-orbit coupling ranges from 1 meV to 300 meV, corresponding to a Rashba spin-splitting of $0.05-10.0$ meV\cite{Varotto2022}, or materials such as InSe, MoSe$_2$, and similar two-dimensional compounds, where the strength of Rashba coupling can be adjusted up to $30$ meV via stacking configurations and other external perturbations\cite{Yuan2013,Farooq2022}. The spin texture and the HHG spectrum for this system are presented in panels (b) and (c) of Fig.\ \ref{fig2},
respectively. Here, the HHG spectrum is computed under a monochromatic light pulse linearly-polarized along the $y$ axis, which is orthogonal to the mirror axis. Even-order harmonics exhibit finite amplitude, consistent with the expectation from the spin texture shown in Fig.\ \ref{fig2}(b), which breaks $\hat{\mathcal{C}}_2$. These harmonics eventually vanish when the $\hat{\mathcal{C}}_2$-breaking warping term approaches zero at $\l=0$ (see panel (c) and inset therein). Additionally, we observe that odd- and even-order harmonics are polarized perpendicular and parallel to the mirror axis, i.e., respectively, along the $y$ and $x$ axes. This is in agreement with the HHG selection rules in Eq.\ \pref{mirror} for a pump field oriented perpendicular to a vertical mirror.

\textit{Systems with broken $\hat{\mathcal{C}}_2$ and T-symmetry} --- To demonstrate that even-order harmonics can effectively probe $\hat{\mathcal{C}}_2$-breaking magnetic patterns in systems lacking inversion and TR symmetry, we consider a square crystal that is intrinsically $\hat{\mathcal{C}}_2$-symmetric but exhibits a specific antiferromagnetic spin arrangement and charge imbalance within the sublattice (Fig.\ \ref{fig2}(d)). Together, these explicitly break the twofold rotational symmetry. The Hamiltonian reads
\be
\hat{H}_0=
\e_\bk^\mu\hat{\tau}_\mu\otimes\hat{\s}_0+
h_\bk^{\mu i}
\hat{\tau}_\mu\otimes\hat{\s}_i,
\ee
where $\hat{\tau}_\mu$ are Pauli matrices representing sub-lattice degrees of freedom. To ensure $\hat{\mathcal{C}}_2$ symmetry breaking, we account for a finite charge imbalance $\D\e$ within the sub-lattice, $\e_\bk^z=\D\e$, and an antiferromagnetic pattern with a local magnetic field $B$ oriented along the $\hat{y}$ direction, $h_\bk^{zy}=B$, as illustrated in Fig.\ \ref{fig2}(d). We also include Rashba coupling, with $h_\bk^{xx}=-\g_R\sin k_y$ and
$h_\bk^{yx}=\g_R\sin k_x$. The resulting spin textures breaks rotational invariance, as clear from Fig.\ \ref{fig2}(e), and lead to non-zero even-order harmonics, as demonstrated in Fig.\ \ref{fig2}(f). These harmonics eventually vanish in the limit of zero charge imbalance, as indicated in the inset of Fig.\ \ref{fig2}(f), since the antiferromagnetic pattern alone is $\hat{\mathcal{C}}_2$-invariant. As a concluding remark, we address the role of Berry curvature in enabling even-order harmonics when TR symmetry is broken. In this case, the spin texture may remain invariant under $\hat{\mathcal{C}}_2$, thereby suppressing even-order harmonics even though the Berry curvature is finite. An example of this scenario is a 2D ferromagnet with Rashba coupling and spin alignment along the out-of-plane direction. This case illustrates that, in the absence of TR symmetry, the rotational symmetry of the spin texture is the key factor in the emission of even-order harmonics.

\begin{figure}[t!]
    \centering
\includegraphics[width=0.42\textwidth]{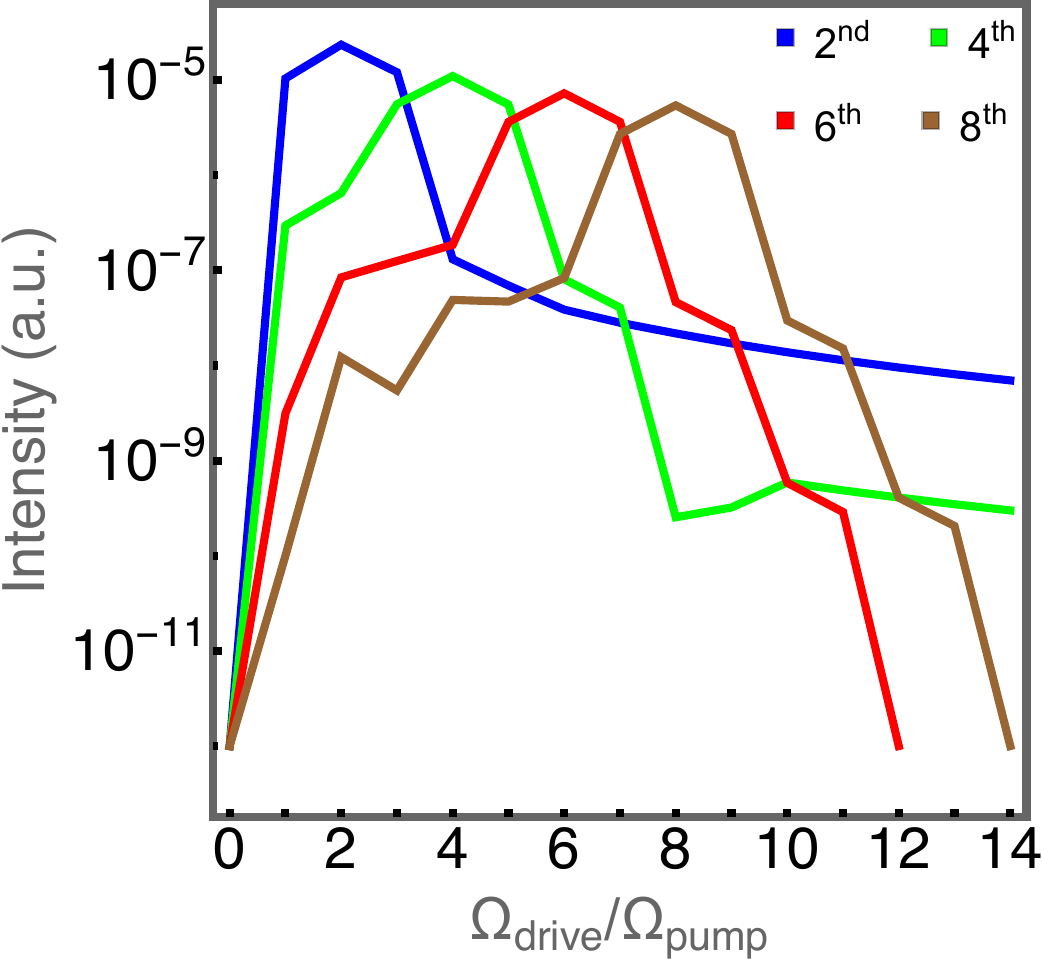}
    \caption{{\bf Dependence of even-order harmonics on the driving frequency.} Amplitudes of the second- (blue), fourth- (green), sixth- (red) and eighth-order harmonics are shown as functions of the ratio between the driving frequency and the pump frequency. The parameters are set to $\Omega_{pump}=0.1$, $\g_R=0.02$, $B_z=0.15$, $B_x=0.1$ and $\mu=0.9$.}
    \label{fig3}
\end{figure}

\textit{Dynamically-broken $\hat{\mathcal{C}}_2$-symmetry} ---So far, we have studied how spin textures and Berry curvature cooperate to enable even-order harmonics when twofold rotational symmetry is broken \textit{homogeneously} in time --that is, through \textit{static} symmetry breaking. We now briefly discuss the case in which rotational symmetry is broken \textit{dynamically} --i.e., when the symmetry-breaking term oscillates in time. This can occur in materials driven into out-of-equilibrium, time-dependent states, which may then exhibit fewer symmetries than the equilibrium ground state. In such cases, rotational symmetry is broken instantaneously but preserved on average, implying that it can be detected only using a time-resolved technique. HHG, in particular, may offer a preferential route, since if the dynamical symmetry breaking occurs on certain timescales, it can only be accessed via the higher-order response of the material, while, for example, the second-order response may be negligible. To further illustrate this point, we consider a system on a square lattice with Rashba SOC and a static magnetic field along the z-axis at equilibrium, and a time-dependent field along $x$ that oscillates in time. The Hamiltonian takes the same form as Eq.\ \pref{HC3}, with 
\bea
h_\mathbf{k}^{(x)}=-\gamma_R\sin k_y+h_\mathbf{k}^{(x)}(t),
\text{ }
h_\mathbf{k}^{(y)}=\gamma_R\sin k_x,
\text{ }
h_\mathbf{k}^{(y)}=B_z.\nn\\
\eea
When $h_\mathbf{k}^{(x)}(t)=0$, the system preserves $\hat{\mathcal{C}}_2$ symmetry. However, introducing a time-dependent magnetic field along $x$, i.e., $h_\bk^{(x)}(t)=B_x\sin\left(\O_{drive}t\right)$, with $\Omega_{drive}$ chosen, for simplicity, as an integer multiple of $\O_{pump}$, breaks $\hat{\mathcal{C}}_2$ symmetry dynamically. We then compute the HHG spectrum for different values of $\Omega_{drive}$. In Fig.\ \ref{fig3}, we show the dependence of the amplitude of the first four even-order harmonics on $\Omega_{drive}$. At $\Omega_{drive}=0$, which corresponds to the absence of the $\hat{\mathcal{C}}_2$-breaking term, all even-order harmonics are zero. As the driving frequency increases, their amplitude grows, with the second-order harmonic being dominant only around $\Omega_{drive}\sim 2\Omega_{pump}$, and higher-order harmonics eventually surpassing it at larger values of $\Omega_{drive}$. This highlights how, for example, second-harmonic generation (SHG) may fail to reveal information about the underlying dynamical $\hat{\mathcal{C}}_2$-symmetry breaking when $\Omega_{drive}$ significantly exceeds $2\Omega_{pump}$. Indeed, when the time scale $\sim 1/\Omega_{drive}$ associated with the dynamical symmetry breaking becomes faster than $\sim 1/(2 \Omega_{pump})$, which is associated with the second-order harmonic, the latter is suppressed due to temporal averaging. Additionally, we observe a correlation between the driving field frequency and the intensity of the generated even harmonics. Specifically, each harmonic's amplitude is modulated by the frequency of the symmetry-breaking field, with harmonics of order $n$ enhanced when $n\Omega_{pump} \sim \Omega_{drive}$. On the other hand, at very large $\Omega_{drive}$, that correspond to the DC limit of the pump pulse, even-order harmonics vanish as the rapidly oscillating $\hat{\mathcal{C}}_2$-breaking term averages to zero over time. This leads, e.g., to a negligible second-order harmonic response or, in the language of nonlinear Hall experiments, a negligible nonlinear second-order Hall current. This particular example underscores the unique ability of high-order harmonics to detect dynamically broken rotational symmetries that are absent at equilibrium and may be inaccessible to conventional probes such as SHG or the nonlinear Hall effect, thereby adding significant value to our previous analysis of HHG selection rules.

\section{Conclusions}

We have investigated the interplay between spin textures, Berry curvature, and HHG in two-dimensional non-centrosymmetric systems. Our analysis reveals that the presence or absence of even-order harmonics is closely tied to the symmetries of the spin textures. Specifically, we established that spin textures with broken twofold rotational symmetry--whether arising from static
or dynamical mechanisms--are essential for the emission of even-order harmonics. Furthermore, we demonstrated the crucial role of Berry curvature in enabling even-order harmonics in systems with time-reversal symmetry. To further validate our general findings, we also provided explicit computations on microscopic models lacking twofold rotational symmetry--both with preserved and broken time-reversal symmetry. In this analysis, we neglected scattering effects and assumed an infinite relaxation time. While we do not expect such effects to qualitatively alter our results, we acknowledge that they nonetheless play a crucial role in shaping realistic harmonic spectra\cite{Vampa2014,Vampa2015,Luu2016}, and that they should be taken into account in future works.

Going back to our results, we emphasize that rotational symmetry can be broken across various degrees of freedom, all linked to the high-harmonic response through spin-orbit interaction. This means that even-order harmonics can identify not only crystalline and spin patterns lacking rotational symmetry but also emergent electronic orderings with spatial reconstructions, such as in kagome systems, where trimerization breaks six-fold rotational symmetry\cite{Zhang2023}. Moreover, nontrivial high-harmonic response can also occur in systems exhibiting specific patterns of spin \cite{Wang2010,Werake2010} and orbitally polarized currents. Our results can also be applied to investigate spin ordering in low-dimensional altermagnets \cite{jeong2024}, marked, for specific terminations, by surface spin textures with odd-parity multipole patterns that break twofold rotational symmetry. All these rotational symmetry breakings, and the associated phase transitions, can be observed through the shift from zero to finite even-order harmonic amplitudes, or conversely, when twofold symmetry is restored from a lower symmetry state. Furthermore, we have demonstrated that HHG is capable of probing dynamical rotational-symmetry breaking in spin textures, which can be utilized to investigate various dynamical spin polarization processes involving a reduction of the rotational symmetry of the spin texture that
arise from phenomena such as light-induced ultrafast magnetization, spin-torque effects, or
charge-to-spin and orbital-to-spin conversion mechanisms. These findings can have, e.g., a potential impact for applications in ultrafast spintronics.

\noindent{\bf{Acknowledgements}}\\
We acknowledge valuable discussions with Jeroen van den Brink. F.F. and F.G. acknowledge support from the Italian Ministry of University and Research (MUR) under Grant PRIN No. 2020JZ5N9M (CONQUEST) and under the National Recovery and Resilience Plan (NRRP), Call PRIN 2022, funded by the European Union - NextGenerationEU, Mission 4, Component 2, Grant No. 2022TWZ9NR (STIMO)-CUP B53D23004560006. M.C. acknowledges partial support from the EU Horizon 2020 research and innovation program under Grant Agreement No. 964398 (SUPERGATE) and, together with F.F., from
NRRP MUR project PE0000023-NQSTI. C.O. acknowledges support by the Italian Ministry of Foreign Affairs and International Cooperation, Grant No. PGR12351 (UTLRAQMAT) and the NRRP MUR project PE0000023-NQSTI (TOPQIN).

\appendix

\section{Eigenstates, density matrix and optical current of a time-periodic quantum system}

In this section, we outline the computation of the optical current for a quantum system under a periodic driving, such as a monochromatic light pulse. For compactness, we set $\hbar=1$ from here on. We begin with an overview of how to compute the eigenvalues and eigenstates of such a system, described by the Hamiltonian $\hat{H}(t)=\hat{H}_0+\th\left(t-t_0\right)\hat{V}(t)$, where $\hat{H}_0$ is the unperturbed Hamiltonian and $\hat{V}(t+T)=\hat{V}(t)$ is a continuous periodic perturbation with period $T=(2\pi)/\O$, switched on at time $t=t_0$. For $t<t_0$, where $\hat{H}(t)=\hat{H}_0$ and the system is undriven, the general solution to the Schr\"{o}dinger Eq. associated with the Hamiltonian can be expressed in the basis of the stationary eigenvectors $\ket{\psi_\mu^0}$ of $\hH_0$ (with the quantum number $\mu$ representing, e.g., momentum and spin) as $\ket{\psi(t)}=\sum_\mu \exp(-i E_\mu^0 (t-t_0))\ket{\psi_\mu^0}$, where $E_\mu^0$ are the unperturbed energy eigenvalues. At $t\geq t_0$, the periodic drive is switched on, and 
\be
\hat{H}(t+T)=\hat{H}(t).
\lb{FL}
\ee
According to the Floquet Theorem\cite{grifoni_physrep98}, a solution to the time-dependent Schr\"{o}dinger Eq. associated with such a time-periodic Hamiltonian can always be written in the form
\be
\ket{\psi_\a(t)}=
e^{-i\e_\a(t-t_0)}
\ket{u_\a(t)},
\lb{floqst}
\ee
with $\a$ a suitable quantum number. $\e_\a$ and $\ket{u_\a(t)}$ are, respectively, the \textit{quasi-energies} and the \textit{Floquet modes}, which satisfy the \textit{quasi-stationary} Schr\"{o}dinger Eq.
\be
\left[\hat{H}(t)-i\pd_t\right]
\ket{u_\a(t)}=\e_\a\ket{u_\a(t)}.
\lb{stat}
\ee
These Floquet modes are periodic in time, and the quasi-energies are defined modulo-$\O$, i.e.,
\be
\ket{u_\a(t+T)}=\ket{u_\a(t)},\quad
-\frac{\O}{2}\leq
\e_\a<\frac{\O}{2}.
\ee
Owing to their time-periodicity, Floquet modes can be expanded in Fourier series as
\be
\ket{u_\a(t)}=
\sum_l   
e^{-il\O t}\ket{u_\a^{(l)}},
\ee
with coefficients $
\ket{u_\a^{(l)}}=\frac{1}{T}
\int_{-T/2}^{T/2}dt 
e^{i l\O t}\ket{u_\a(t)}$. In Fourier space, the quasi-stationary Schr\"{o}dinger Eq. \pref{stat} becomes a time-independent eigenvalue equation, i.e., $\left(\hat{H}_{mn}- m\O\d_{mn}\right)\ket{u_\a^{(n)}}=\e_\a\ket{u_\a^{(m)}}$, with $\hat{H}_{mn}=\frac{1}{T}
\int_{-T/2}^{T/2}dt 
e^{i (m-n)\O t}\hat{H}(t)$ the Fourier transform of the Hamiltonian. This equation can be solved using standard eigenvalue algorithms. The full evolution of the system can be then summarized by the density matrix $\hat{\rho}(t)$, which inherits the time-periodicity of the Hamiltonian. This implies that $\hat{\rho}(t)$ must be diagonal in the basis of the Floquet modes, i.e.,
\be
\hr(t)=\sum_\a
\rho_{\a}\ket{u_\a(t)}\bra{u_\a(t)},
\lb{rho}
\ee
with time-independent diagonal matrix elements, in the limit $t_0\ra -\infty$, given by
\be
\rho_{\a}=\sum_{l,\mu}
\abs{\braket{u_\a^{l}}{\psi_\mu^0}}^2
f\left( E_\mu^0 \right),
\ee
with $f(x)\equiv \left(1+\exp\left(\b x\right)\right)^{-1}$ the Fermi function at the inverse temperature $\b=1/T$. Having computed the eigenstates and the density matrix of the driven system, the optical current can be obtained as the quantum thermal average $\bJ(t)=\Tr\left[\hat{\rho}(t)\hat{\bv}(t)
\right]$ of the velocity operator $\hat{\bv}(t)=1/i\left[\hat{H}(t),\hat{\br}\right]$. $\bJ$ can be expanded in Fourier series, and using Eq.\ \pref{rho}, we can write the $l$-th Fourier component of $\bJ$, which is responsible for the emission of the $l$-th-order harmonic, as
\bea
\bJ_l =\sum_\a \rho_\a
\sum_{mn} \bra{u_\a^{(m)}}
\hat{\bv}_{m+l-n}
\ket{u_\a^{(n)}},
\eea
where $\hat{\bv}_{l}\equiv\frac{1}{T}
\int_{-T/2}^{T/2}dt\hat{\bv}(t)
e^{i l\O t}$ is the $l$-th Fourier component of the velocity operator.

\section{Spin textures and HHG selection rules}

In this section we derive the selection rules for the spin textures and the high-order harmonics. Starting from the former, these are computed as the quantum thermal average $\boldsymbol{\s}_\bk\equiv\tr\left[\hat{\rho}_\bk^0\hat{\boldsymbol{\s}}_\bk\right]$, with $\hat{\rho}_\bk^0=\sum_{\bk,s}f\left(E_{\bk s}^0\right)\ket{\psi_{\bk s}^0}\bra{\psi_{\bk s}^0}$ the density matrix of the undriven system. In the absence of external perturbations, a system of electrons confined in a periodic potential remains invariant under the action of a space-group symmetry, i.e., a combination of a point-group symmetry $\hat{\mathcal{P}}$, such as a rotation or a mirror, and a space translation $\hat{\tau}$ within a primitive cell. For simplicity, and as previously done in the manuscript, we refer to a generic space-group symmetry $\lbrace{\hat{\mathcal{P}}|\hat{\tau}\rbrace}$ using only its point-group element $\hat{\mathcal{P}}$. Denoting by $P$ the matrix associated with the action of $\hat{\mathcal{P}}$ on a 2D polar vector, the spin, that transforms as a pseudovector, obeys $\boldsymbol{\s}_{P\bk}=\pm\det\left(P\right)P\boldsymbol{\s}_{\bk}$, for $\hat{\mathcal{P}}$ unitary/anti-unitary. For example, for the twofold rotational symmetry $\hat{\mathcal{C}}_2$ we have that $P=-\mathbb{1}$ is minus the 2x2 identity, causing the out-of-plane and the in-plane spin components to transform as described by Eq.\ (4) of the main text. 

To derive the selection rules for HHG, we introduce the concept of \textit{dynamical symmetry}\cite{janssen_phys69,alon_prl98,neufeld_natcomm19,lysne_prb20}. When a time-dependent external driving field $\bE(t)=-\pd_t\bA(t)$, where $\bA(t)$ is the associated vector potential in the Weyl gauge, is introduced via the minimal-coupling substitution of the bare momentum $\bp$ with $\tilde{\bp}\equiv\bp+\bA(t)$\cite{mandlshaw}, the resulting time-dependent Hamiltonian $\hat{H}(t)$ generally loses its invariance under $\hat{\mathcal{P}}$. However, it may exhibit invariance under a dynamical symmetry $\hat{g}\equiv\hat{\mathcal{P}}\otimes\hat{\mathcal{T}}$, which combines $\hat{\mathcal{P}}$ with the unitary operator $\hat{\mathcal{T}}$ acting on time. For the following discussion, we consider a monochromatic drive $\bA(t)=\Re\lbrace \bA_0\exp(i\O t)\rbrace$ oscillating with frequency $\O$, where $\bA_0$ is the polarization vector with, e.g., $\bA_0\propto \left(\cos\th,\sin\th,0\right)$ for linearly-polarized light, with $\th$ the polarization angle, or $\bA_0\propto (2)^{-1/2}\left(1,\pm i,0\right)$ for left-hand/right-hand circularly-polarized light. In this case, the time symmetry is a time-translation operator $\hat{\mathcal{T}}_{T/N}\equiv\exp\left(i\left(T/N)\right)\pd_t\right)$, which shifts time by a fraction of the pulse period $T\equiv 2\pi/\O$, with $N$ an integer. This operator must ensure that the canonical momentum $\tilde{\bp}$ transforms under $\hat{g}$ in the same way that the bare momentum $\bp$ transforms under $\hat{\mathcal{P}}$. We consider once again the case of $\hat{\mathcal{C}}_2$ as an example. Since bare momentum transforms as $\hat{\mathcal{C}}_2\bp\hat{\mathcal{C}}_2^{\dagger}=-\bp$, the time-dependent case requires $\hat{\mathcal{T}}_{T/2}$, which shifts time by half a period and transforms the vector potential as $\hat{\mathcal{T}}_{T/2}\bA(t)\hat{\mathcal{T}}_{T/2}^{\dagger}=-\bA(t)$. Consequently, the canonical momentum transforms as $\hat{g}\tilde{\bp}\hat{g}^{\dagger}=-\tilde{\bp}$ under $\hat{g}\equiv\hat{\mathcal{T}}_{T/2}\otimes\hat{\mathcal{C}}_2$, leaving the time-dependent Hamiltonian invariant and leading to the selection rule discussed in Eq.\ (3) of the main text. As we will see in the following, not all point-group symmetries have a dynamical counterpart.

The invariance of $\hat{H}(t)$ under $\hat{g}$ imposes a constraint on the quantum thermal averages of physical observables. Specifically, we focus on the optical current $\bJ$. Since $\left[\hat{g},\hat{H}(t)\right]=0$, the density matrix, which evolves in time according to the Liouville-von Neumann Eq.\ $i\pd_t\hat{\rho}(t)=[\hat{H}(t),\hat{\rho}(t)]$, transforms as $\hat{g}\hr(t)\hat{g}^{\dagger}=
\hr(\pm t)$, depending on whether $\hat{\mathcal{P}}$, and consequently $\hat{g}$, is unitary/antiunitary. In the time domain, this implies that $\Tr\left[\hat{\rho}(t)\hat{\bv}(t)
\right]=\Tr\left[\hat{\rho}(\pm t)\hat{g}\hat{\bv}
(t)\hat{g}^{\dagger}
\right]$. From this identity, one can derive the following selection rule for $\bJ_n$ at any order $n$ in Fourier space:
\be
P\bJ_n=e^{i \frac{2\pi n}{N}}\bJ_n^{(*)},
\lb{SR}
\ee
Eq.\ \pref{SR} applies to (anti-)unitary point-group symmetries and, upon defining $\tau\equiv 2\pi n/N$, it is equivalent to Eq.\ (2) of the main text. Notably, if $\bJ_n^*=\bJ_n$, with either $\Re\bJ_n=0$ or $\Im\bJ_n=0$, the $n$-th order harmonic is linearly-polarized. Conversely, if $\abs{\Re\bJ_n}=\abs{\Im\bJ_n}$ and $\Re\bJ_n\cdot\Im\bJ_n=0$, it is circularly-polarized. In the following, we apply Eq. \pref{SR} to specific cases of 2D point-group symmetries relevant for the main text, i.e., rotations around the $z$ axis, vertical mirrors, and time reversal. Some of the following selection rules have also been discussed in Ref.\ \cite{neufeld_natcomm19} in the context of 3D bulk systems.

\textit{Rotations} A rotation $\hat{\mathcal{C}}_z(\phi)\equiv\hat{R}_z(\phi)\otimes\hat{U}_z(\phi)$ by an angle $\phi$ around the vertical $z$-axis combines a 2D real-space rotation $\hat{R}_z(\phi)$ with a corresponding spin rotation $\hat{U}_z(\phi)$. The bare momentum transforms as $\hat{\mathcal{C}}_z(\phi)\bp\hat{\mathcal{C}}_z^{\dagger}(\phi)=Q(\phi)\bp$, where $Q(\phi)$ is the rank-2 rotation matrix. For $\phi=\pi$, i.e., the case of the twofold rotational symmetry $\hat{\mathcal{C}}_z(\pi)\equiv\hat{\mathcal{C}}_2$, one defines, as already seen above, the dynamical symmetry $\hat{g}\equiv\hat{\mathcal{T}}_{T/2}\otimes\hat{\mathcal{C}}_2$ for the time-dependent Hamiltonian, both in the case of linearly-polarized and circularly-polarized light. Eq.\ \pref{SR} thus yields $-\bJ_n
=\exp(i n\pi)\bJ_n$ for the current, i.e., a finite $\bJ_n$ for $n$ odd, with no restriction on its polarization, and a vanishing $\bJ_n$ for even $n$. In other words, the optical response of systems invariant under $\hat{\mathcal{C}}_2$ lacks even harmonics. For $\phi=2\pi/N$, with $N=3,4$ or $6$, the rotation can be mimicked only in the case of circularly-polarized external light by choosing the time-translation operator as $\hat{\mathcal{T}}_{T/N}$, since $\hat{\mathcal{T}}_{T/N}\bA(t)\hat{\mathcal{T}}_{T/N}^{\dagger}=\hat{Q}(2\pi/N)\bA(t)$. With this choice of $\hat{g}$ Eq.\ \pref{SR} yields $\hat{Q}(2\pi n/N)\bJ_n=\exp(i 2\pi/N)\bJ_ n$, as also discussed in Ref. \cite{alon_prl98}, which only allows harmonics with circular polarization $(1,\pm i, 0)$ and order $n=Nq\pm 1$, with $q$ an integer.

\textit{Mirrors} Under a vertical mirror $\hat{\mathcal{M}}_v(\mathbf{u})$, where $\mathbf{u}$ denotes its in-plane normal direction, the bare momentum transforms as $\hat{\mathcal{M}}_v(\mathbf{u})\bp\hat{\mathcal{M}}_v(\mathbf{u})^{\dagger}=2(\mathbf{u}\cdot\bp)\mathbf{u}-\bp$. While circularly-polarized light always breaks mirror symmetries, $\hat{g}\equiv\hat{\mathcal{T}}_{T/2}\otimes\hat{\mathcal{M}}_v(\mathbf{u})$ is a dynamical symmetry for a linearly-polarized pump with $\bA\perp\mathbf{u}$. In this case Eq.\ \pref{SR} requires $\bJ_n$ to be linearly-polarized, with $\bJ_n\pll\mathbf{u}$ for even $n$ and $\bJ_n\perp\mathbf{u}$ for odd $n$. Consequently, for a pump field perpendicular to a mirror, even/odd harmonics are linearly-polarized parallel/perpendicularly to its normal direction. When the pump field is polarized along the mirror ($\bA\pll\mathbf{u}$) the dynamical symmetry is just $\hat{g}\equiv\hat{\mathcal{M}}_v(\mathbf{u})$. Along with Eq.\ \pref{SR}, this requires an optical response polarized linearly in the direction of the mirror ($\bJ_n\pll\mathbf{u}$) for all $n$.

\textit{Time reversal} If the unperturbed Hamiltonian is time-reversal invariant and the pump-field is linearly-polarized (circularly-polarized light always breaks time reversal) the anti-unitary operator $\hat{g}\equiv\hat{\mathcal{T}}_{T/2}\otimes\hat{\mathrm{T}}$ involving the time reversal operator $\hat{\mathrm{T}}$ is a dynamical symmetry of the time-dependent Hamiltonian. In this case, Eq.\ \pref{SR} requires $\bJ_n=\pm\bJ_{-n}$ for odd/even harmonics, implying that the optical response must be linearly polarized at any order.

\textit{Product of two symmetry elements} By combining two symmetry elements, further selection rules can be derived. The two relevant cases are the anti-symmetry versions of rotations and vertical mirrors, given respectively by the two products $\hat{\mathrm{T}}\otimes C_z(\phi)$ and $\hat{\mathrm{T}}\otimes\hat{\mathcal{M}}_v(\mathbf{u})$. In the first case we have that only for $\phi=\pi$ and a linearly-polarized pump field there is a dynamical symmetry $\hat{g}\equiv\hat{\mathrm{T}}\otimes C_z(\pi)$, that acts as an identity on the momentum and the velocity and thus implies $\bJ_n=\bJ_{-n}$ for all $n$, with no restrictions on the allowed harmonic orders. This is the case, e.g., of the antiferromagnetic systems investigated in the main text. In the case of a mirror combined with the time reversal the dynamical symmetry is $\hat{g}\equiv \hat{\mathrm{T}}\otimes\hat{\mathcal{M}}_v(\mathbf{u})$ for $\bA\perp\mathbf{u}$, and the selection rule dictates $\bJ_n\perp \mathbf{u}$ for all $n$. Conversely, when $\bA\pll\mathbf{u}$ the dynamical symmetry is $\hat{g}\equiv \hat{\mathcal{T}}_{T/2}\otimes\hat{\mathrm{T}}\otimes\hat{\mathcal{M}}_v(\mathbf{u})$, and it must be $\bJ_n\perp\mathbf{u}$ for even $n$, $\bJ_n\pll\mathbf{u}$ for odd $n$.

\section{Time-evolution and current operator in the Bloch basis}

In the previous sections and throughout most of this paper, we performed calculations in the Floquet basis and using the length gauge. While we emphasize that the results, in particular the HHG spectrum, are independent of the basis and gauge-invariant, selecting a specific combination can be convenient for addressing certain problems. For analyzing the influence of the Berry curvature on the HHG spectrum, the most natural natural choice is the Bloch eigenstate basis, labeled by the crystal momentum $\bk$ and by the band index $\m$, and it is convenenient to work in the length gauge, in which the time-dependent Hamiltonian in the presence of an external field $\bE(t)$ takes the form (with $e=1$ for simplicity)
\be
\hH(t)=\hH_0-\hat{\br}\cdot \bE(t),
\ee
where the position operator $\hat{\br}$ reads, in the Bloch basis,
\be
\bra{\psi^0_{\bk \mu}}\hat{\br}\ket{\psi^0_{\mathbf{k'} \n}}=
i\d_{\m\n}\bdnb_\bk\d\left(\bk-\mathbf{k'}\right)+
\d\left(\bk-\mathbf{k'}\right)\bxi_{\bk\m\n}
\ee
In this representation, the band resolved current matrix elements are
\bea
\bJ_{\bk\m\n}(t)&=&\frac{1}{i}\bra{\psi^0_{\bk\m}}
\left[\hat{\br},\hat{H_0}\right]\ket{\psi^0_{\bk\n}}\nn\\
&-&\frac{1}{i}\cdot\bra{\psi^0_{\bk\m}}
\left[\hat{\br},\bE(t)
\cdot\hat{\br}\right]
\ket{\psi^0_{\mathbf{k}\n}}.
\lb{Jmn}
\eea
The first contribution incorporates the intraband and the interband currents and can be obtained by standard computation as\cite{aversa_prb95}
\be
\frac{1}{i}\bra{\psi^0_{\bk\m}}
\left[\hat{\br},\hat{H_0}\right]\ket{\psi^0_{\bk\n}}=
\bdnb_\bk\e_{\bk\n}\d_{\m\n}+
i(\e_{\bk\m}-\e_{\bk\n})\bxi_{\bk\m\n}.
\lb{contr1}
\ee
The second contribution is the anomalous one, and it can be computed by noting that\cite{aversa_prb95}
\bea
\bra{\psi^0_{\bk\m}}
\left[\hat{\br}_\a,\hat{\br}_\b\right]
\ket{\psi^0_{\mathbf{k}\n}}
&=& i\left(
\pd_{k_\a}\bxi_{\bk\m\n}^\b-
\pd_{k_\b}\bxi_{\bk\m\n}^\a \right.\nn\\
&&\left.-i\sum_\l
\left(
\bxi_{\bk\m\l}^\a \bxi_{\bk\l\n}^\b-
\bxi_{\bk\m\l}^\b \bxi_{\bk\l\n}^\a
\right)\right)\nn\\
&=&i\bO_{\bk mn}^{\a\b},
\lb{contr2}
\eea
\\
i.e., the Berry Curvature. Plugging Eqs.\ \pref{contr1} and \pref{contr2} into Eq.\ \pref{Jmn} and performing the thermal average, one exactly obtains Eq.\ \pref{bJ} of the main text.

In a similar fashion, the Boltzmann Eq.\ \pref{boltz} for the density-matrix elements, $\hat{\rho}_{\bk\m\n}(t)\equiv\bra{\psi^0_{\bk\m}}
\hat{\rho}(t)\ket{\psi^0_{\mathbf{k}\n}}$, can be obtained by expressing the Liouville-von Neumann equation, $d\hat{\rho}(t)/dt=-i\left[\hH_0,\hat{\rho}(t)\right]+i\bE(t)\cdot[\hat{\br},\hat{\rho}(t)]$, in the Bloch representation and noting that
\bea
\bra{\psi^0_{\bk\m}}\left[\hat{\br},\hat{\rho}(t)\right]
\ket{\psi^0_{\mathbf{k}\n}}&=&
i\bdnb_\bk\rho_{\bk\m\n}(t)\nn\\
&+&
\sum_l\left(
\bxi_{\bk\m\l}\rho_{\bk\l\n}(t)-
\rho_{\bk\m\l}(t)\bxi_{\bk\l\n}
\right)\nn\\
\eea

\bibliography{Literature.bib}

\end{document}